\documentclass[12pt]{article}
\usepackage{epsfig,psfig,graphicx,rotating}
\def\be{\begin{equation}}
\def\ee{\end{equation}}
\begin{document}
\centerline{\bf STATISTICAL MECHANICS OF THE SELF-GRAVITATING GAS:}
\centerline{\bf THERMODYNAMIC LIMIT, UNSTABILITIES AND  PHASE DIAGRAMS}
\begin{center}
{\bf H. J. de Vega$^{(b,a)}$}
\footnote{devega@lpthe.jussieu.fr}
{\bf N. G. Sanchez$^{(a)}$}
\footnote{Norma.Sanchez@obspm.fr}

\bigskip

{$^{(a)}$ Observatoire de Paris, LERMA, Laboratoire Associ\'e au 
CNRS \\ UMR 8112, 61, Avenue de l'Observatoire, 75014 Paris, France. }

\bigskip

{$^{(b)}$ LPTHE, Laboratoire Associ\'e au CNRS UMR 7589,\\
Universit\'e Pierre et Marie Curie (Paris VI) et Denis Diderot (Paris VII),\\
Tour 24, 5 \`eme. \'etage, 4, Place Jussieu, 75252 Paris, Cedex 05,
France.}
\end{center}
\centerline{\bf Abstract}
We show that the self-gravitating gas at thermal equilibrium has an infinite 
volume limit in the three ensembles (GCE, CE, MCE) when 
$ (N, V)  \to \infty $, keeping $ N/ V^{1/3}$ fixed, that is, with
$ \eta \equiv {G \, m^2 N \over V^{1/3} \; T} $ 
fixed. We develop Monte Carlo simulations, analytic mean field methods (MF) 
and low density expansions. We {\bf compute} the equation of state and find 
it to be {\bf locally } $ p(\vec r) = T \, \rho_V(\vec r) $, that is a 
{\bf local ideal gas}  equation of state.
The system is in a gaseous phase for $ \eta < \eta_T  = 1.51024 \ldots$ and
collapses into a very dense object for $ \eta > \eta_T $ in the 
CE with the pressure
becoming large and negative. The isothermal compressibility diverges at 
$ \eta = \eta_T $. We compute the fluctuations around mean field
for the three ensembles.  We show that the particle distribution 
can be described by a Haussdorf dimension $ 1 < D < 3$.

Nous montrons que la limite de volume infini existe pour le gaz auto-gravitant 
\`a l'equilibre thermique dans les trois ensembles (EGC,EC,EMC) quand 
$ (N, V)  \to \infty $, avec $ N/ V^{1/3}$ fixe, c'est \`a dire 
$ \eta \equiv {G \, m^2 N \over V^{1/3} \; T} $ fixe. Nous utilisons les
simulations Monte Carlo, la m\'ethode du champ moyen et les developpements
\`a basse densit\'e. Nous calculons l'\'equation d'\'etat et nous 
trouvons qu'elle
est {\bf localement}  $ p(\vec r) = T \, \rho_V(\vec r) $, c'est à dire, 
l'équation d'un gaz parfait {\bf local}.
Le system est dans une phase gazeuse pour $ \eta < \eta_T  = 1.51024 \ldots$ et
s'effondre dans un objet tr\`es dense pour $ \eta > \eta_T $ dans l'ensemble
canonique avec une pression grande et n\'egative. 
La compressibilit\'e isothermique
diverge \`a $ \eta = \eta_T $. Nous calculons les fluctuations autour du champ
moyen pour les trois ensembles. Nous montrons que 
la distribution des particules
est d\'ecrite par une dimension de Haussdorf  $ 1 < D < 3$.
\date{\today}
\bigskip

Physical systems at thermal equilibrium are usually homogeneous. This is the 
case for gases with short range  intermolecular forces (and in absence
of external fields). In such cases the entropy is maximum when the
system homogenizes.

When long range interactions as the gravitational force are present, even
the ground state is inhomogeneous. In this case,  each element of the
substance is acted on by very strong forces due to distant
particles in the gas. Hence, regions near to and far from the boundary of the 
volume occupied by the gas will be in very different conditions, and, as a 
result, the homogeneity of the gas is destroyed \cite{llms}. The state
of maximal entropy for gravitational systems is {\bf inhomogeneous}. 
This  basic  inhomogeneity suggested us that fractal
structures can arise in a self-interacting gravitational
gas \cite{natu,I,II,pal,sieb}. 

The inhomogeneous character of the ground state for gravitational
systems explains why the universe is {\bf not} going towards a `thermal
death'. A `thermal death' would mean that the universe evolves towards
more and more homogeneity. This can only happen if the entropy is
maximal for an homogeneous state. Instead, it is the opposite what
happens, structures are formed in the universe through
the action of the gravitational forces as time evolves.

Usual theorems in statistical mechanics break down for inhomogeneous
ground states. For example, the specific heat may be negative in the
microcanonical ensemble (not in the canonical ensemble where it is
always positive)\cite{llms}. 

As is known, the thermodynamic limit for self-gravitating systems does
not exist in its usual form ($N\to \infty,\; V \to \infty,\; N/V = $
fixed). The system collapses into a very dense phase which is
determined by the short distance (non-gravitational) forces between
the particles. However, the thermodynamic functions {\bf exist} in the 
{\bf dilute} limit \cite{I,II,pal}
$$ 
N\to \infty\; ,\; V \to \infty\; ,\; {N\over V^{1/3}} = \mbox{fixed} \; ,
$$ 
where $ V $ stands for the volume of the box containing the gas.
In such a limit, the energy $E$, the free energy and the entropy turns to be
extensive. That is, we find that they take the form of $ N $ times a
function of the intensive dimensionless variables:
$$
\eta = {G \, m^2 N \over L \; T} \quad \mbox{or} \quad
\xi = { E \, L \over G \, m^2 \, N^2}
$$
where $\eta$ and $\xi$ are  intensive variables. Namely, $\eta$ and
$\xi$ stay finite when $ N $ and $ V \equiv L^3 $ tend to infinite. The variable $\eta$ is
appropriate for the canonical ensemble and $\xi$ for the
microcanonical ensemble. Physical magnitudes as the specific heat,
speed of sound, chemical potential and  compressibility only depend on
$\eta$ or $\xi$. The variables $\eta$ and $\xi$, as well as the ratio $ N/L $, are
therefore  {\bf intensive} magnitudes. 
The energy, the free energy, the 
Gibbs free energy and the entropy are of the form $ N $ times a
function of $\eta$. These functions of $\eta$ have a finite $ N = \infty $
limit for fixed $\eta$ (once the ideal gas contributions are
subtracted). Moreover, the dependence on $ \eta $ in all these
magnitudes express through a single universal function $ f(\eta) $. 
The variable $\eta$ is the ratio of the characteristic gravitational energy
$\frac{G m^2 N}{L}$ and the kinetic energy $T$ of a particle in the gas. 
For $\eta=0$ the ideal gas is recovered.

In refs. \cite{I,II} and \cite{pal} we have thoroughly studied
the statistical mechanics of the self-gravitating
gas. That is,  our starting point is the partition function for
non-relativistic particles interacting through their gravitational
attraction in thermal equilibrium. We study the self-gravitating gas in the
three ensembles: microcanonical (MCE), canonical (CE) and grand
canonical (GCE). We performed calculations by three methods:

\begin{itemize}

\item{By expanding
the partition function through direct calculation in powers of $1/\xi$
and $\eta$ for the MCE and CE, respectively.  These expressions apply
in the dilute regime ($ \xi \gg 1 \, , \, \eta \ll 1 $) and become
identical for both ensembles for $ N \to \infty $. At $ \eta = 0 = 1/\xi$ 
we recover the ideal gas behaviour.}

\item{By performing Monte Carlo simulations both in the MCE
and in the CE.  
We found in this way that the self-gravitating gas {\bf collapses} at a
critical point which depends on the ensemble considered. As shown in
fig. \ref{fig1} the collapse occurs first in the canonical ensemble (point
T). The microcanonical ensemble exhibits a larger region of stability
that ends at the point MC (fig.  \ref{fig1}). Notice that the
physical magnitudes are identical in the common region of validity of
both ensembles within the statistical error. Beyond the critical point
T the system 
becomes suddenly extremely compact with a large negative pressure in
the CE. Beyond the point MC in the MCE the pressure and the
temperature increase suddenly and the gas collapses. 
The {\bf phase transitions} at T and at MC are of {\bf zeroth order} since
the Gibbs free energy has discontinuities in both cases.}

\begin{figure}
\begin{turn}{-90}
\resizebox{10cm}{!}{\includegraphics{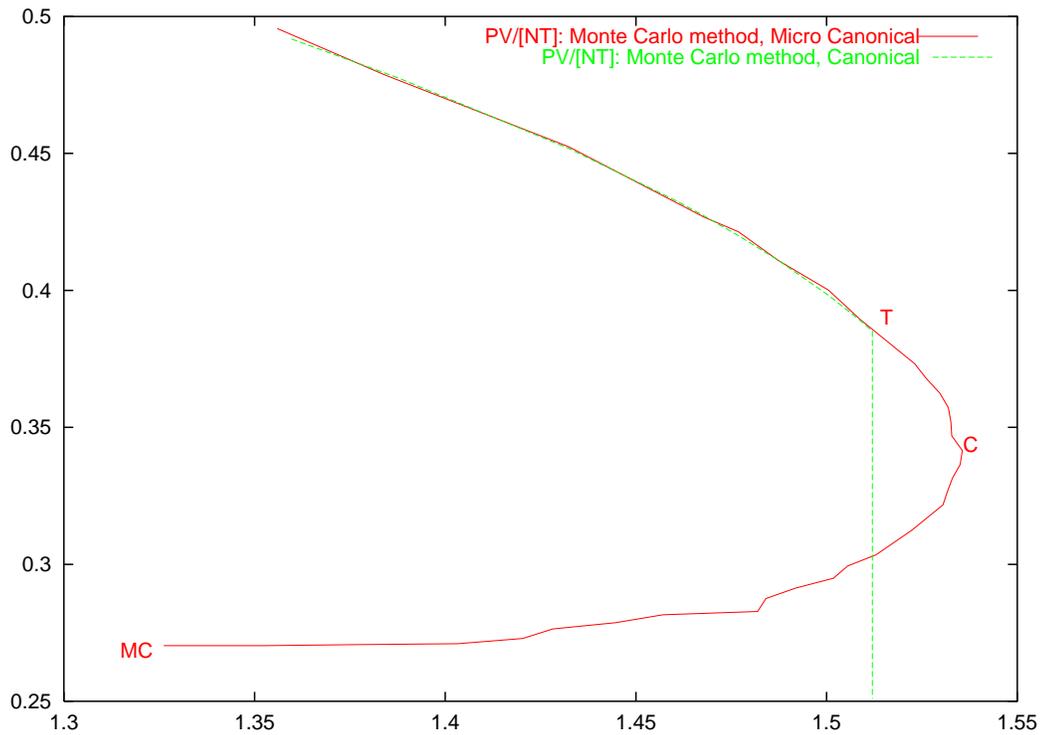}}
\end{turn}
\caption{ $f(\eta) = P V/[ N T]$ as a function of $ \eta $  by Monte
Carlo simulations for the microcanonical and canonical ensembles
($N=2000$). Both curves coincide within the statistical error till the point T.
\label{fig1}}
\end{figure}

\item{By using the mean field  approach we evaluate the partition
function for large $ N $. We do this computation  in the grand canonical,
canonical and microcanonical ensembles. In the three cases, the
partition function is expressed as a functional integral over a
statistical weight which depends on the (continuous) particle
density. These statistical weights are of the form of the exponential
of an `effective action' proportional to $ N $. Therefore, the $ N \to
\infty $ limit follows by the saddle point method. The saddle point is
a space dependent mean field showing the inhomogeneous character of the ground
state. Corrections to the mean field are of the order $ 1/N $ and can
be safely ignored for $ N \gg 1 $ except near the critical
points. These mean field results turned out to be in  excellent agreement
with the Monte Carlo results and with the low density expansion. }
\end{itemize}
We calculate the saddle point (mean field)  for spherical symmetry
and we obtain from it the various physical
magnitudes (pressure, energy, entropy, free energy, specific heats,
compressibilities, speed of sound and particle density).
Furthermore, we computed in ref.\cite{II} the {\bf determinants of small
fluctuations} around the saddle point solution for spherical symmetry
for the three statistical ensembles.

When any small fluctuation around the saddle point decreases the
statistical weight in the functional integral, the saddle point 
dominates the functional integral and the mean field approach can be valid. 
In that case, the  determinant of small fluctuations is positive. A
negative determinant of small fluctuations indicates that some
fluctuations around the saddle point are increasing the statistical
weight in the functional integral and hence the saddle point {\bf does
not} dominate the partition function. The mean field approach cannot be
used when the determinant of small fluctuations is negative. We find analytically
in the CE that the determinant of small fluctuations vanishes at the point $\eta_C$ = 1.561764 and becames negative at $\eta > \eta_C$ \cite{I}. (The point $\eta_C$ is indicated $C$ in Figure 1). We find that the CE specific heat $c_V$ at the point $C$ diverges as  \cite{I}:
$$
c_V \buildrel{ \eta \uparrow \eta_C}\over= \pm 0.63572\ldots
(\eta_C-\eta)^{-1/2} - 0.19924\ldots+ {O}(\sqrt{\eta_C-\eta})
$$
The (+) sign refers to the positive (first) branch, and the (-) sign to the negative (second) branch (between the points $C$ and $MC$). 

However, it must be noticed that the {\bf instability point} is located at $\eta =\eta_T < \eta_C$, as shown by both, mean field and Monte Carlo computations. (The point $\eta_T$ is indicated $T$ in Figure 1).
The onset of instability in the canonical ensemble
coincides with the point where the isothermal compressibility diverges. 
The isothermal compressibility $\kappa_T$ is positive from $\eta=0$ 
till $\eta=\eta_T=1.51024 \ldots$. At this point $\kappa_T$ as well
as the specific heat at constant pressure $c_P$ diverge and change
their signs \cite{I}. Moreover, at this point 
the speed of sound at the center of the sphere becomes imaginary \cite{II}. 
Therefore, small density fluctuations will {\bf grow} exponentially 
in time instead of exhibiting 
oscillatory propagation. Such a behaviour leads to the collapse of the gas into
a extremely compact object.
Monte-Carlo simulations confirm the presence of this instability at 
$ \eta_T = 1.510 \ldots$ in the canonical ensemble and the formation
of the collapsed object \cite{I}. 

The collapse in the Grand Canonical ensemble (GCE) occurs for a smaller 
value of $ \eta = \eta_{GC} = 0.49465\ldots $, while in the microcanonical
ensemble, the collapse arrives later, in the second sheet, at 
$ \eta =  \eta_{MC} = 1.25984\ldots$.

We find that the Monte Carlo simulations for self-gravitating gas in
the CE and the MCE confirm the stability results obtained from mean field. 

The saddle point solution is {\bf identical} for the three statistical
ensembles. This is not the case for the fluctuations around it. The
presence of constraints in the CE (on the number of particles) and in
the MCE (on the energy and the number of particles) {\bf changes} the
functional integral over the quadratic fluctuations with respect to
the GCE. 

The saddle point of the partition function turns out to coincide with the
hydrostatic treatment of the self-gravitating gas \cite{sasbt} (which is 
usually known as the `isothermal sphere' in the spherically symmetric case).

We find that the {\bf Monte Carlo} simulations (describing thermal equilibrium)
are much more {\bf efficient} than the $N$-body simulations integrating 
Newton's equations of motion. (Indeed, the integration of Newton's equations
provides much more detailed information than the one needed in thermal 
equilibrium investigations).
Actually, a few hundreds of particles are enough to get
quite accurate results in the Monte Carlo simulations (except near the collapse
points). Moreover, the Monte Carlo results turns to be in {\bf excellent}
agreement with the mean field calculations up to very small corrections 
of the order $ (1/N) $. 
Our Monte Carlo simulations are performed in a cubic geometry. The
equilibrium configurations obtained in this manner can thus be called
the {\bf `isothermal cube'}.

In summary, the picture we get from 
our calculations using these three  methods show that the self-gravitating gas
behaves as a perfect gas for  $ \eta \to 0, \; 1/\xi\to 0 $. When $
\eta $ and $ 1/\xi $ grow, the gas becomes denser till it suddenly
condenses into a high density object at a critical point GC, C or MC
depending upon the statistical ensemble chosen. 

$ \eta $ is related with the
Jeans' length $ d_J $ of the gas through $ \eta = 3 \; ( L/d_J )^2
$. Hence, when $ \eta $ goes beyond $ \eta_T $, the length of the
system becomes larger than $ d_J / \sqrt{\eta_T /3} $. 
The collapse at
T in the CE is therefore a manifestation of the Jeans' instability. 

\begin{figure}[t] 
\begin{turn}{-90}
\resizebox{10cm}{!}{\includegraphics{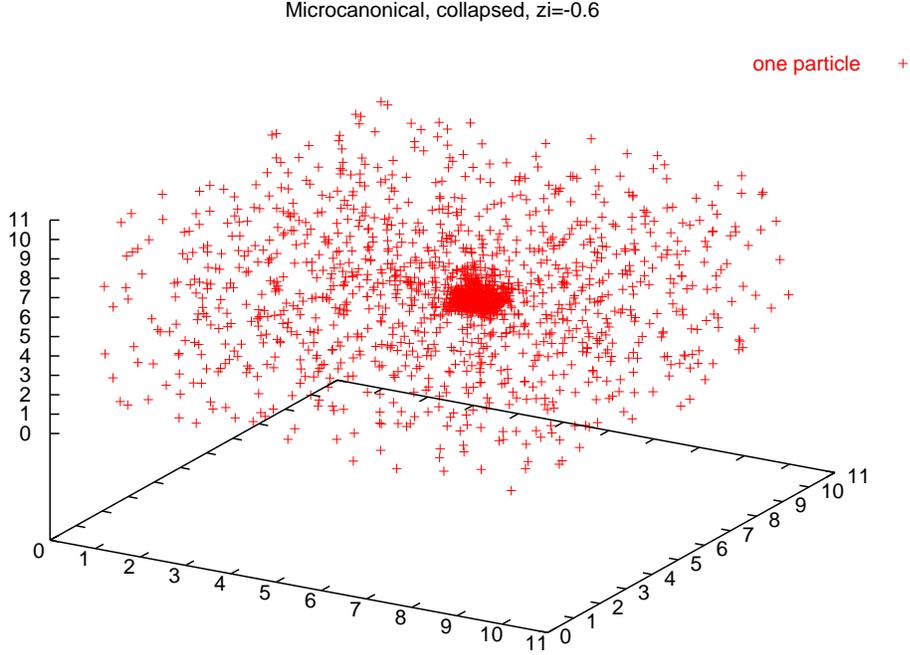}}
\end{turn}
\caption{Average particle distribution in
the collapsed phase from Monte Carlo simulations with $2000$ particles
in the microcanonical ensemble for $\xi = - 0.6, \; \eta = 0.43, \;
pV/[NT] = 0.414 $.  \label{colmc}}  
\end{figure}

In the MCE, the determinant of fluctuations vanishes at the point MC.
The physical states
beyond MC are collapsed configurations as shown by the Monte Carlo
simulations (see fig. \ref{colmc}). Actually, the gas collapses in the
Monte Carlo simulations slightly before the mean field prediction for
the point MC. The phase transition at the microcanonical critical point MC is
the so called gravothermal catastrophe \cite{lynbell2}.

The gravitational interaction being attractive without lower
bound, a short distance cut-off ($ A $) must be introduced in order
to give a meaning to the partition function. We take the gravitational
force between particles as $ -G \; m^2 /r^2 $ for $ r > A $ and zero
for $ r< A $ 
where $ r $ is the distance between the two particles. We show that
the cut-off effects are negligible in the $ N = \infty $ limit. That
is, {\bf once} we set $ N = \infty $ with fixed $ \eta $,
{\bf all} physical quantities are {\bf finite} in the
zero cut-off limit ($ A = 0$). The cut-off effects are of the order $
A^2/L^2 $ and can be safely ignored.

In ref. \cite{I} we expressed all global physical quantities in terms of a
single function  $f(\eta)$. Besides
computing numerically  $ f(\eta) $ in the mean field approach,
we showed that this function obeys a
first order non-linear differential equation of first Abel's type.
We obtained analytic results about  $ f(\eta) $  from the {\bf Abel's equation}. 
$f(\eta)$ exhibits a {\bf square-root cut} at  $\eta_{C}$, the critical
point in the CE. The first Riemann sheet is realized both
in the CE and the MCE, whereas the second  Riemann sheet (where $ c_V
< 0 $) is only realized in the MCE. $ f(\eta) $ has infinitely
many branches in the $\eta$ plane but only the first two branches are
physically realized. Beyond MC the states described by the mean field
saddle point are unstable. 

We plot and analyze the equation of state, the
energy, the entropy, the free energy, $ c_V $ and the isothermal
compressibility (figs. 9-13 in ref.\cite{I}).
Most of these physical magnitudes were not previously
computed in the literature as functions of $ \eta $.

We find analytically the behaviour of $f(\eta)$ near the point $ \eta_C $ in
mean field,
$$
f(\eta) \buildrel{ \eta \uparrow \eta_C}\over= \frac13 +
0.27137\ldots \sqrt{\eta_C-\eta} +0.27763\ldots\; (\eta_C-\eta)
+ O\left[(\eta_C-\eta)^{3/2}\right] \; .
$$
This exhibits a square root {\bf branch point} singularity at $\eta_C$. This shows that the specific heat at constant volume diverges  at $\eta_C$ as
$\left(\eta_C-\eta \right)^{-1/2} $ for $ \eta \uparrow
\eta_C$. However,  it must be noticed, that the specific heat at constant pressure and the isothermal
compressibility? both  diverge at the point $ \eta_T $ as $ \left(\eta_T-\eta
\right)^{-1} $. These mean field results apply for $ |\eta-\eta_C| \ll
1 \ll N|\eta-\eta_C| $. Fluctuations around mean field can be
neglected in such a regime.

The Monte Carlo calculations permit us to obtain  $f(\eta)$ in the
collapsed phase. Such result cannot be
obtained in the mean field approach. The mean field approach only provides
information as $f(\eta)$ in the dilute gas phase. 

For the self-gravitating gas, we find that the Gibbs free energy $ \Phi
$ {\bf is not} equal to $ N $ times the chemical potential and that the
thermodynamic potential $ \Omega $ {\bf is not} equal to $ - PV $ as
usual \cite{llms}. This is a consequence of the dilute thermodynamic
limit $ N \to \infty, \; L\to \infty, \; N/L=$fixed.  

We computed the determinant of small fluctuations 
around the saddle point solution for spherical symmetry in all three
statistical ensembles. 
In the spherically symmetric case, the determinant of small
fluctuations is written as an infinite product over partial waves. The
S and P wave determinants are written in closed form in terms of the
saddle solution. The determinants for higher partial waves are
computed numerically. All partial 
wave determinants are positive definite except for the S-wave \cite{II}. 
The reason why the fluctuations are different in the three ensembles
is rather simple. The more contraints are imposed the smaller becomes
the space of fluctuations. Therefore, in the grand canonical ensemble
(GCE) the system is more free to fluctuate and the phase transition
takes place earlier than in the micro-canonical (MCE) and canonical
ensembles (CE). For the same reason, the transition takes place
earlier in the CE than in the MCE.

The conclusion being that  the mean field correctly gives an excellent
description of the thermodynamic limit except near the critical points
(where the small fluctuations determinant vanishes);
the  mean field is valid for $N|\eta-\eta_{crit}|\gg 1$. The vicinity of the
critical point should be studied in a double scaling limit $N \to
\infty,\; \eta \to \eta_{crit}$. Critical exponents are reported in
ref. \cite{I} for $ \eta \to \eta_C $ using the mean field. These mean field
results apply for $ |\eta-\eta_C| \ll 1 \ll N|\eta-\eta_C| $ with $ N
\gg 1$. Fluctuations around mean field can be neglected in such a regime.

We computed {\bf local} properties of the gas in \cite{II}. That is, the
local energy density $ \epsilon(r) $, local particle density, local
pressure and the local speed of sound. Furthermore, 
we analyze the scaling behaviour of the particle
distribution and its fractal (Haussdorf) dimension \cite{II}.

The particle distribution $\rho_V({\vec q})$ proves to be {\bf inhomogeneous}
(except for $\eta \ll 1$) and described by an universal function of $\eta$,
the geometry and the ratio ${\vec r} = {\vec q} / R, \; R$ being the
radial size. Both Monte Carlo simulations and the Mean Field approach
show that the system is inhomogeneous forming a clump of size smaller than the box
of volume $ V $ [see fig. \ref{colmc} here and figs. 3, 5 and 6 in \cite{I}].
 
The particle density in the bulk behaves as $ \rho_V({\vec q}) \simeq r^{D-3} $. 
That is, the mass $ M(R) 
$ enclosed on a region of size $ R $ vary approximately as
$$
 M(R) \simeq C \;  R^D \; . 
$$
$ D $ slowly decreases from the value $ D = 3 $ for the ideal gas
($\eta=0$) till $ D = 0.98 $ in the extreme limit of the MC point,
$D$ takes the value $1.6$ at $\eta_C$,  [see Table 1]. 
This indicates the presence of a fractal distribution with
Haussdorf dimension $D$. 

Our study of the statistical mechanics of a self-gravitating system indicates 
that gravity provides a dynamical mechanism to produce fractal
structures \cite{natu,I,II}.

The average distance between particles monotonically decrease with $
\eta $ in the first sheet. The mean field and Monte Carlo are very
close in the gaseous phase whereas the Monte Carlo simulations exhibit
a spectacular drop in the average particle distance at the clumping
transition point T. In the second sheet (only
described by the MCE) the average particle distance increases with $
\eta $ \cite{II}. 

We find that the {\bf local} equation of state is given by 
$$
 p(\vec r) = T \, \rho_V(\vec r) \; .
$$
\noindent We have  {\bf derived} the equation of state for the
self-gravitating gas. It is {\bf locally} the {\bf ideal gas
} equation, but the self-gravitating gas being inhomogeneous,  the
pressure at the surface of a given volume is not equal
to the temperature times the average density of particles in the volume.
In particular, for the whole volume: $ PV/[NT] = f(\eta) \leq 1 $
(the equality holds only  for $ \eta = 0 $).  

Notice that we have found 
the local ideal gas equation of state $ p(\vec r) = T \, \rho_V(\vec r) $ 
for purely gravitational interaction between
particles. Therefore, equations of state different from this one, (as often
assumed and used in the literature for the self-gravitating gas), necessarily imply the presence of additional non-gravitational forces. 

The local energy density $\epsilon(r)$ turns out to be an increasing function of $r$ in the
spherically symmetric case. The energy density is always positive on
the surface, whereas it is positive at the center for $ 0 \leq \eta  <
\eta_3  = 1.07783\ldots $, and 
negative beyond the point $ \eta = \eta_3  = 1.07783\ldots $.

The local {\bf speed of sound} $ v_s^2(r) $ is computed in the mean field approach as a
function of the position for spherical symmetry and long wavelengths. 
$ v_s^2(r) $ diverges at $ \eta  =  \eta_T  = 1.51024\ldots $ in the
first Riemann sheet. Just beyond this point $ v_s^2(r) $ is large and
{\bf negative} in the bulk showing the strongly unstable behaviour of
the gas for such range of values of $ \eta $. 

Moreover, we have
shown the {\bf equivalence} between the statistical mechanical
treatment in the mean field approach and the hydrostatic description
of the self-gravitating gas \cite{sasbt}.

The success of the hydrodynamical description depends on the value of
the mean free path ($l$) compared with the relevant sizes in the system. 
$ l $ must be $ \ll 1 $. 
We compute   the ratio $ l/a $ ({\bf Knudsen} number), where $ a $ is a
length scale that stays fixed for $ N \to \infty $
and show that  $ l/a \sim N^{-2} $. This
result ensures the accuracy of the hydrodynamical description for
large $ N $. 

Furthermore, we have computed in ref. \cite{II} several
physical magnitudes as functions of $ \eta $ and $ r $ which were not
previously 
computed in the literature as the speed of sound, the energy density,
the average distance between particles and we notice the presence of a
{\bf Haussdorf} dimension in the particle distribution. 

The statistical mechanics of a selfgravitating gas formed by
particles with different masses is thoroughly investigated in
ref.\cite{sieb} while the selfgravitating gas in the presence of 
the cosmological constant is thoroughly investigated
in refs. \cite{negra}. In ref.\cite{cluster} the Mayer expansion
for the selfgravitating gas is investigated in connection with
the stability of the gaseous phase.

\bigskip

\begin{tabular}{|l|l|l|}\hline
$ \eta $ & $\hspace{0.5cm} D $ & $\hspace{0.5cm} C $\\ \hline 
$ 0.06204 $ &    \hspace{0.3cm} $ 2.97 $ \hspace{0.3cm}  &  \hspace{0.3cm}
$ 1.0 $ \hspace{0.3cm}  \\ \hline 
$ \eta_{GC} = 0.49465\ldots $ & \hspace{0.3cm} $ 2.75  $\hspace{0.3cm} &
\hspace{0.3cm} $ 1.03 $ \hspace{0.3cm} \\ \hline 
 $ 1.24070 $  & \hspace{0.3cm} $ 2.22 $ \hspace{0.3cm} &\hspace{0.3cm} $ 1.1
$\hspace{0.3cm} \\ \hline 
$ \eta_{C} = 1.561764\ldots $ & \hspace{0.3cm} $ 1.60
$\hspace{0.3cm} & \hspace{0.3cm} 
$ 1.07 \hspace{0.3cm} $ \\ \hline
$ \eta_{MC} = 1.25984\ldots $ & \hspace{0.3cm} $ 0.98  $
\hspace{0.3cm}&\hspace{0.3cm} 
$ 1.11 $\hspace{0.3cm} \\ \hline 
\end{tabular}

\bigskip

{TABLE 1. The Fractal Dimension $ D $ and the proportionality
coefficient $ C $ as a function of $ \eta $ from a fit to the mean
field results according to $  M (r) \simeq C \; r^D $.}
The point $ \eta_{MC}$ is on the second Riemann sheet.

\end{document}